\documentclass[aps, prd, twocolumn, showpacs, superscriptaddress, groupedaddress]{revtex4}
\usepackage{graphicx}	
\usepackage{amssymb}
\usepackage{dcolumn}
\usepackage{tcolorbox}
\usepackage{color}
\usepackage{mathtools}
\usepackage{subfigure, rotating, bm, array}
\usepackage{hyperref}
\usepackage{caption}

\begin{document}
\title{Energy extraction–driven instability and horizon formation in Kerr–Newman naked singularities and their limiting cases}

\author{Vishva Patel}
\email{vishvapatelnature@gmail.com}
\affiliation{PDPIAS, Charusat University, Anand 388421, India}
\affiliation{ International Centre for Space and Cosmology, School of Arts and Sciences, Ahmedabad University, Ahmedabad 380009, India }

\date{\today}

\begin{abstract}

Energy extraction from compact objects has been a central topic in general relativity since the introduction of the Penrose process. In this work we present a unified analysis of rotational and electromagnetic energy extraction in Kerr, Reissner–Nordström, and Kerr–Newman spacetimes. Using particle energetics and the irreducible mass formalism, we compare the efficiencies of these mechanisms and examine their consequences for horizonless objects. While purely rotational extraction in Kerr spacetime is fundamentally limited by geometric constraints, electromagnetic interactions increase the region of negative energy orbits through an effective ergoregion, allowing significantly higher efficiencies. In Kerr–Newman geometry, the combined effect of rotation and charge further enhances the extractable energy. We then investigated the long-term evolution of over-extremal cases under continuous extraction. By deriving coupled evolution equations for the mass, spin, and charge parameters, we show that continuous extraction can gradually drive a naked singularity toward the extremal bound. For astrophysically realistic luminosities, the characteristic evolution timescale is of order $\sim 10^{9}$ years. These results suggest that energy extraction provides an energetic indication of instability in Reissner–Nordström, Kerr, and Kerr–Newman naked singularities and may lead to horizon formation as a long-term stabilizing outcome.

\bigskip
\textbf{Key words:} Penrose process, Magnetic Penrose process, Energy extraction, Naked singularities.
\end{abstract}

\maketitle

\section{Introduction}
\label{sec:intro}

The extraction of energy from black holes stands as one of the most striking ideas in general relativity. Roger Penrose and R. M. Floyd first showed that the spacetime geometry of a rotating black hole makes it possible to extract its rotational energy \cite{PenroseFloyd:1971}. In the Kerr metric, an ergoregion exists where particles can follow trajectories with negative conserved energy as seen by an observer at infinity. In the ergosphere of a rotating black hole, an incoming particle can split into two fragments: one that plunges into the black hole carrying negative energy (as measured from a observer at infinity), while the other escapes with greater energy than the original particle possessed. This mechanism is called the Penrose process. Yet the Penrose process has clear practical limits. Wald’s careful analysis revealed kinematic constraints that demand highly tuned trajectories and relativistic relative speeds between fragments, resulting in extremely low realistic efficiencies \cite{Wald:1974kya}. These requirements make it unlikely to explain powerful astrophysical sources like quasars or relativistic jets, prompting researchers to seek more viable mechanisms in magnetized environments.\\ 

During the early 1970s, Denardo and Ruffini explored an analogous energy extraction mechanism applicable to charged Reissner–Nordström spacetimes \cite{DenardoRuffini:1973}. Building on Christodoulou’s introduction of the irreducible mass concept, they established a framework for quantifying how much energy could, in principle, be stored in the electromagnetic field of a charged black hole and extracted even without rotation \cite{Christodoulou1970}. Soon afterward, De Felice examined the instability of Kerr naked singularities and the potential astrophysical implications of rotational energy extraction in the Kerr geometry \cite{deFelice:1978}.\\

A major step forward came with the Blandford–Znajek mechanism \cite{BlandfordZnajek:1977}, an electromagnetic extension of Penrose process. Here, a rotating black hole threaded by large scale magnetic fields generates an electromotive potential that drives currents and extracts energy through the magnetosphere, ultimately converted into Poynting flux and synchrotron radiation. This mechanism provides a self-consistent explanation for relativistic jets in active galactic nuclei (AGN) and X-ray binaries. Later studies confirmed that electromagnetic energy extraction dominates in realistic astrophysical scenarios compared to the Penrose process \cite{Livio1998}.\\

In 1985, Bhat, Dhurandhar, and Dadhich investigated the Penrose process in the Kerr–Newman geometry \cite{Bhat1985}. They showed that charge enlarges the region of negative energy orbits and give higher extractable energy, bridging the Kerr and Reissner–Nordström limits even without external fields. Wagh and Dadhich later extended these ideas to black holes in external electromagnetic fields, demonstrating substantial efficiency gains from charge–magnetic coupling \cite{Wagh1985,Wagh1989}. Magnetic fields soon emerged as central rather than secondary in black hole energetics. The Magnetic Penrose Process (MPP), revisited and formalized by Tursunov and Dadhich \cite{Tursunov2019}, revealed that charged particles in magnetized rotating spacetimes can access negative energy orbits beyond the geometric ergoregion. The Lorentz force relaxes the fine-tuning needed in the original process, giving efficiencies that often exceed 100\% and surpass classical limits by orders of magnitude. Which offers a robust foundation for high-energy astrophysical sources.\\

Alongside MPP developments, particle acceleration and collisional studies gained momentum after Bañados, Silk, and West showed that particles can reach arbitrarily high center-of-mass energies near extremal Kerr horizons, known as BSW effect \cite{PhysRevLett.103.111102}. Patil and Joshi extended this analysis to Kerr, Reissner–Nordström, and naked singularity spacetimes, showing that ultra-energetic collisions can achieve high center-of-mass energies without the fine-tuning required near black hole horizons \cite{Patil2014a,Patil2014b,Patil2012}. After that work by Grib \& Pavlov, Mukherjee \& Nayak, and others connected these collisional Penrose variants to jet production and observable high-energy emission \cite{Grib2014,Mukherjee2018}. Modern research continues to build on these foundations. Komissarov revisited the relative roles of Blandford–Znajek and Penrose processes, underscoring electromagnetic dominance in magnetohydrodynamic accretion regimes \cite{Komissarov2008}. More recent studies have examined collisional Penrose in Kerr naked singularities and energy extraction in magnetized Reissner–Nordström geometries, while comprehensive reviews link modern energy extraction process to jet formation, synchrotron radiation, and high-energy signatures \cite{Mukherjee2018,Shaymatov2022,Stuchlik2021}. Recent advancements in the Penrose process across diverse geometries, such as, regular, quantum-corrected, and accelerating black holes are explored in \cite{
Zeng:2026egq,Chen:2026oxr,Bambhaniya:2025gai,Hejda:2025cfy,Zhao:2025ouq,Kar:2025anc,Vertogradov:2025wgg,Zhang:2025blr,Xamidov:2025pln,Feiteira:2024dwn,Xamidov:2024wou,Viththani:2024map,Shaymatov:2024ibv,Shaymatov:2024fle,Chakraborty:2024aug,Feiteira:2024awb,Zaslavskii:2023jxu,Turimov:2023mia,Kurbonov:2023uyr,Patel:2023efv,Vertogradov:2022eeq,Patel:2022jbk,Ma:2022zow,Zaslavskii:2022vap,Shaymatov:2022eyz,Stuchlik:2021unj,Tursunov:2021jjf,Hejda:2021cbk,Gupta:2021vww,Zaslavskii:2020fmz,Tursunov:2019oiq,Zaslavskii:2018kix,Dadhich:2018gmh,Liu:2018myg,Vicente:2018mxl,Toshmatov:2014qja,Ganguly:2014pwa,Lasota:2013kia,Dadhich:2012yu,Liu:2012qe,Abdujabbarov:2011af,Prabhu:2009ju,Komissarov:2008yh,Grib:2014nha,Peng:2023zvf,Mukherjee:2018cbu,Armaza:2015eha,Pugliese:2018hju,Pugliese:2021ivl,Nozawa:2005eu,Livio:1998qi,Khodadi:2022dff,Khan:2019gco,Ruiz:2012te,Chen:2025bph}, while progress in collisional Penrose processes and their connections to particle acceleration is detailed in  \cite{Chatterjee:2026uxg,
Liberati:2021uom,Kokubu:2020jvd,Zaslavskii:2020dts,Coimbra-Araujo:2020awz,Zaslavskii:2019bho,Zaslavskii:2015vaa,Patil:2011ya,Patil:2011uf,Patil:2011aa,Turimov:2025tmf}.\\

It has been over half a century since the mechanism of rotational energy extraction from black holes was first proposed. Over the years, extensive research has explored both rotational and electromagnetic extraction processes, in black holes as well as horizonless compact objects. In this work we present a unified framework for rotational and electromagnetic energy extraction across the foundational geometries of Reissner–Nordström, Kerr, and Kerr–Newman spacetimes, and examine their implications for the stability of naked singularities. In particular, we show that continuous energy extraction can drive over-extremal configurations toward the extremal bound, leading to horizon formation on long astrophysical timescales.\\

The organization of the paper is as follows. In Sec.~\ref{sec:kn} we review the Kerr–Newman spacetime and define the extremality parameter that distinguishes black holes from naked singularities. Section~\ref{sec:PP_MPP} briefly summarizes the Penrose process and the magnetic Penrose process, emphasizing the roles of rotational and electromagnetic energy extraction. In Sec.~\ref{sec:framework} we introduce the general framework for Penrose-type energy extraction and derive the coupled evolution equations governing the mass, spin, and charge parameters. Sections~\ref{sec:kerr_ns} and~\ref{sec:rn_ns} analyze the Kerr and Reissner–Nordström naked singularity limits, where closed form expressions for the horizon formation time are obtained. The general Kerr–Newman case is discussed in Sec.~\ref{sec:kn_ns}. Section~\ref{sec:examples} presents numerical examples and astrophysical timescale estimates. Finally, Sec.~\ref{sec:discussion} summarizes the results and discusses their implications for the long-term evolution and stability of naked singularities. Throughout the paper we adopt the metric signature ($-$,+,+,+) and set $G=c=1$.

\section{Kerr--Newman Spacetime}
\label{sec:kn}

The Kerr--Newman spacetime is the most general stationary, axisymmetric, asymptotically flat solution of the Einstein--Maxwell equations, characterised by mass $M$, spin parameter $a$, and electric charge $Q$.  In Boyer--Lindquist coordinates $(t,r,\theta,\phi)$ the metric is,
\begin{widetext}
\begin{equation}
ds^{2} = -\frac{\Delta_{KN} - a^{2}\sin^{2}\theta}{\Sigma}\,dt^{2}
- \frac{2a\sin^{2}\theta(r^{2}+a^{2}-\Delta_{KN})}{\Sigma}\,dt\,d\phi
+ \frac{(r^{2}+a^{2})^{2}-\Delta_{KN} a^{2}\sin^{2}\theta}{\Sigma}
  \sin^{2}\theta\,d\phi^{2}
+ \frac{\Sigma}{\Delta_{KN}}\,dr^{2} + \Sigma\,d\theta^{2},
\label{eq:KNmetric}
\end{equation}
\end{widetext}
where
\begin{equation}
\Delta_{KN} = r^{2}-2Mr+a^{2}+Q^{2},\qquad
\Sigma = r^{2}+a^{2}\cos^{2}\theta.
\label{eq:DeltaSigma}
\end{equation}
The associated electromagnetic four-potential is
\begin{equation}
A_{\mu}dx^{\mu} = -\frac{Qr}{\Sigma}\,(dt-a\sin^{2}\theta\,d\phi).
\label{eq:fourpotential}
\end{equation}
Horizons occur where $\Delta_{KN} = 0$, giving the outer (event) horizon
$r_{+}$ and the inner (Cauchy) horizon $r_{-}$,
\begin{equation}
r_{\pm} = M \pm \sqrt{M^{2}-a^{2}-Q^{2}}.
\label{eq:horizons}
\end{equation}
For $M^{2}>a^{2}+Q^{2}$ both horizons are real and the spacetime describes a black hole.  The extremal case $M^{2}=a^{2}+Q^{2}$ merges them at $r_{+}=r_{-}=M$, while $M^{2}<a^{2}+Q^{2}$ gives no real horizons. Hence the spacetime represents a naked singularity, in which the ring singularity at $r=0$ is visible to distant observers, violating cosmic censorship in classical general relativity. It is therefore convenient to define the \emph{extremality parameter} as,
\begin{equation}
\Lambda \;\equiv\; a^{2}+Q^{2}-M^{2},
\label{eq:Lambda}
\end{equation}
so that $\Lambda<0$ (black hole), $\Lambda=0$ (extremal), and $\Lambda>0$ (naked singularity).  The Kerr--Newman geometry smoothly interpolates between Kerr ($Q=0$), Reissner--Nordstr\"{o}m
($a=0$), and Schwarzschild ($a=Q=0$) spacetimes.

\section{Penrose and Magnetic Penrose Processes: A Brief Overview}
\label{sec:PP_MPP}

\subsection{Penrose process}

The Penrose process extracts rotational energy from a spinning black hole via its ergoregion. The region between the event horizon and the static limit surface is known as the ergoregion. Here, the timelike Killing vector becomes spacelike. Within the ergoregion the conserved energy measured at infinity,
\begin{equation}
E = -p_{t},
\label{eq:energy}
\end{equation}
can be negative for certain geodesics. In the standard scenario, an incident particle enters the ergoregion and splits into two fragments. One fragment crosses the event horizon carrying negative energy ($E_{(1)}<0$), while the other escapes to infinity with $E_{(2)}>E_{(0)}$. Energy and angular-momentum conservation require,
\begin{equation}
E_{(0)} = E_{(1)}+E_{(2)},\qquad L_{(0)} = L_{(1)}+L_{(2)},
\label{eq:ELconserv}
\end{equation}
so the energy of the escaping fragment is extracted from the black hole's rotational energy. For an equatorial motion, the four-momentum is
\begin{equation}
P^{\mu} = P^{t}\!\left(\partial_{t}+v\,\partial_{r}
          +\Omega\,\partial_{\phi}\right),
\end{equation}
where $v=dr/dt$ and $\Omega=d\phi/dt$, and the extraction efficiency is defined as,
\begin{equation}
\eta = \frac{E_{(2)}}{E_{(0)}}-1.
\label{eq:eta_penrose}
\end{equation}
The maximum efficiency is achieved when the splitting occurs arbitrarily close to the event horizon with vanishing radial velocities.

\begin{table}[htbp]
\centering
\caption{Global extractable energy in Kerr–Newman black holes and
efficiencies.}
\label{tab:kn_extract}
\begin{tabular}{||c|c|c|c|c|c||}
\hline
$a$ & $Q$ & $r_{+}$ & $M_{\rm ir}$ &
$E_{\rm extr}$ ($M$ units) & Efficiency (\%) \\
\hline
0.1  & 0.1  & 1.98995 & 0.99623 & 0.00376973 & 0.376973 \\
0.5  & 0.5  & 1.70711 & 0.889412 & 0.110588  & 11.0588  \\
0.7  & 0.7  & 1.14142 & 0.669485 & 0.330515  & 33.0515  \\
0.1  & 0.99 & 1.0995  & 0.552018 & 0.447982  & 44.7982  \\
0.5  & 0.7  & 1.5099  & 0.795268 & 0.204732  & 20.4732  \\
0.7  & 0.5  & 1.5099  & 0.832136 & 0.167864  & 16.7864  \\
0.99 & 0.1  & 1.0995  & 0.739763 & 0.260237  & 26.0237  \\
\hline
\end{tabular}
\end{table}

\begin{figure*}[ht!]
\centering
\subfigure[Ergoregion structure for Kerr black hole ($a=0.99$)]
  {\includegraphics[width=5cm]{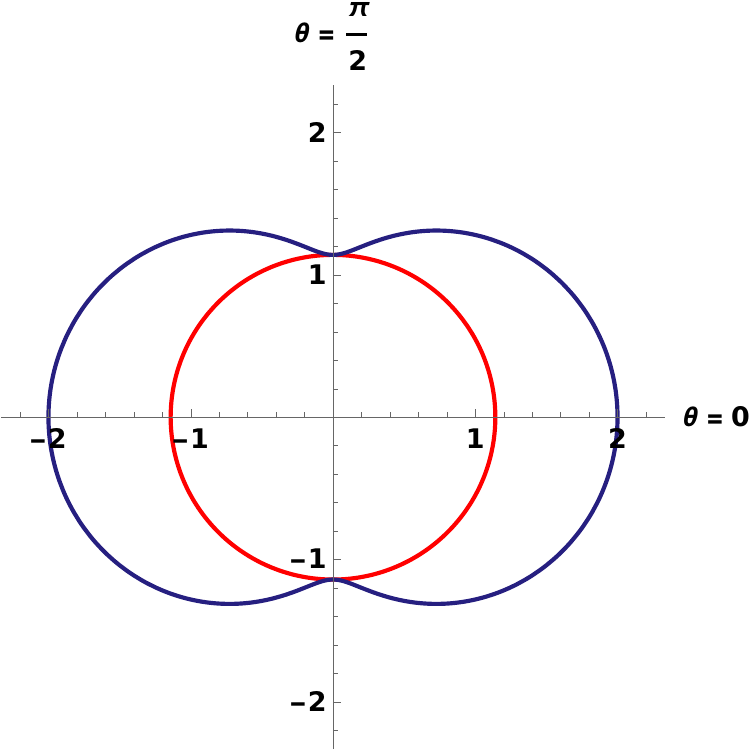}\label{ergokerr}}
\hspace{1cm}
\subfigure[Negative energy orbits for Kerr black hole (various $a$)]
  {\includegraphics[width=8cm]{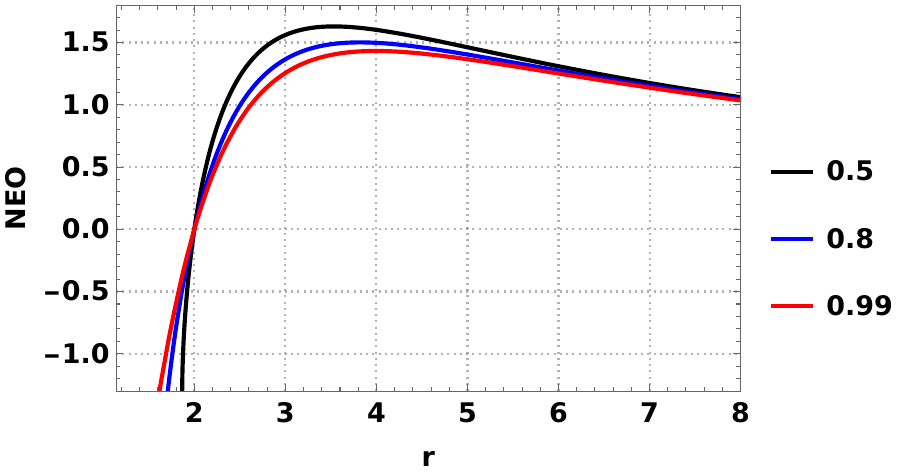}\label{NEOkerr}}
\caption{(a) Ergoregion structure for a Kerr black hole with
spin parameter $a=0.99$.  (b) Negative energy orbits for a Kerr
black hole with different spin parameters.}
\label{fig:kerr}
\end{figure*}

\begin{table}[htbp]
\centering
\begin{minipage}{0.48\textwidth}
\centering
\captionof{table}{Reissner--Nordstr\"{o}m black holes}
\label{tab:RN_Efficiency}
\begin{tabular}{||c|c|c||}
\hline
$Q/M$ & $r_{+}/M$ & Efficiency (\%) \\ \hline
0.10 & 1.99499 & 0.250628 \\
0.30 & 1.95394 & 2.30304  \\
0.50 & 1.86603 & 6.69873  \\
0.70 & 1.71414 & 14.2929  \\
0.90 & 1.43589 & 28.2055  \\
0.99 & 1.14107 & 42.9466  \\ \hline
\end{tabular}
\end{minipage}
\hfill
\begin{minipage}{0.48\textwidth}
\centering
\captionof{table}{Kerr black holes (Penrose process)}
\label{tab:penrose_efficiency_kerr}
\begin{tabular}{||c|c||}
\hline
Spin $a$ & Efficiency $\eta_{\max}$ (\%) \\ \hline
0.10 & 0.062  \\
0.30 & 0.585  \\
0.50 & 1.763  \\
0.70 & 4.008  \\
0.90 & 9.009  \\
0.99 & 16.195 \\ \hline
\end{tabular}
\end{minipage}
\end{table}

\begin{table}[htbp]
\centering
\caption{Estimated maximal energy extraction efficiency in
over-extremal (naked) KN geometry.}
\label{tab:naked_efficiency}
\begin{tabular}{||c|c|c|c||}
\hline
$a$ & $Q$ & $\alpha$ & $\eta_{\max}$ (\%) (MPP) \\ \hline
0.8 & 1.1 & 0.05 & $\sim$120 \\
1.0 & 1.3 & 0.15 & $\sim$250+ \\
0.5 & 1.2 & 0.10 & $>200$ (diverging near $r\to0$) \\ \hline
\end{tabular}
\end{table}

\subsection{Magnetic Penrose process}

The magnetic Penrose process (MPP) extends the original idea by incorporating electromagnetic interactions.  When a rotating compact object is immersed in an external electromagnetic field, the conserved energy of a charged particle becomes,
\begin{equation}
E = -p_{t}-qA_{t},
\label{eq:energy_MPP}
\end{equation}
where $q$ is the particle charge and $A_{t}$ is the electromagnetic potential Eq.~\ref{eq:fourpotential}. The
additional electromagnetic term significantly enlarges the region of negative energy orbits by forming an effective ergoregion beyond the geometric one. As in the classical case, an incident charged particle may split into two fragments: one following a negative energy orbit ($E_{3}<0$) falls into the compact object, while the other escapes to infinity carrying more energy than the original particle. From the Charge conservation we obtain:
\begin{equation}
E_{1} = E_{2}+E_{3},\qquad
q_{1} = q_{2}+q_{3},\qquad
L_{1} = L_{2}+L_{3}.
\label{eq:MPP_conserv}
\end{equation}
The energy of the escaping particle is
\begin{equation}
E_{2} = \chi(E_{1}+q_{1}A_{t})-q_{2}A_{t},
\end{equation}
where
\begin{equation}
\chi = \left( \frac{\Omega_{1}-\Omega_{3}}{\Omega_{2}-\Omega_{3}} \right) \cdot
       \left( \frac{g_{tt}+\Omega_{2}g_{t\phi}}{g_{tt}+\Omega_{1}g_{t\phi}}\right),
\end{equation}
and the MPP efficiency;
\begin{equation}
\eta = \frac{E_{2}-E_{1}}{E_{1}}
\label{eq:eta_MPP}
\end{equation}
can significantly exceed the classical Penrose process limit of $\sim20.7\%$. In the limit of vanishing electromagnetic coupling ($q=0$), MPP reduces smoothly to the classical mechanism.

\subsection{Rotational and electromagnetic energy extraction}
\label{sec:rot_em}

In Kerr spacetime ($Q=0$), energy extraction is purely based on frame-dragging: the ergoregion increases with $a$ (Fig.~\ref{fig:kerr}), yet even at the extremal limit the efficiency is bounded at $\sim20.7\%$ (Table~\ref{tab:penrose_efficiency_kerr}). In the Reissner--Nordstr\"{o}m spacetime ($a=0$) there is no ergoregion, but the conserved energy of a radially moving charged particle,
\begin{equation}
E = m\sqrt{f(r)}-\frac{qQ}{r},\qquad
f(r) = 1-\frac{2M}{r}+\frac{Q^{2}}{r^{2}},
\label{eq:RN_energy}
\end{equation}
becomes negative for oppositely charged particles ($qQ<0$), forming an effective ergoregion. As $Q/M\to1$, a substantial fraction of the total mass-energy becomes extractable (Table~\ref{tab:RN_Efficiency}), with
efficiencies surpassing the purely rotational case.\\

In the general Kerr--Newman case both spin and charge contribute to this process. Using the extremality parameter~$\Lambda$ defined in Eq.~\eqref{eq:Lambda}, Table~\ref{tab:kn_extract} shows that even for $\Lambda<0$ (black hole regime) the combined rotational and electromagnetic effects enhance extractable energy beyond the Kerr limit. For $\Lambda>0$ (naked singularity), the absence of an event horizon altogether removes the irreducible-mass bound: maximal efficiency can grow without limit, as illustrated in Table~\ref{tab:naked_efficiency} using the parameter $\alpha\equiv\sqrt{a^{2}+Q^{2}}/M-1$, which is related to $\Lambda$ by $\Lambda=M^{2}\alpha(\alpha+2)$. The divergence of efficiency with increasing $\alpha$ provides quantitative support for the energetic instability of naked singularities under continuous extraction, which we develop quantitatively in the following sections.

\section{Penrose-Type Energy Extraction: General Framework}
\label{sec:framework}

\subsection{Horizon formation as a stability criterion}

As established in Section~\ref{sec:kn}, horizon existence requires $\Lambda=a^{2}+Q^{2}-M^{2}\leq0$. Continuous energy extraction that drives $\Lambda$ from a positive value to zero would signal the formation of an event horizon and the conservation of cosmic censorship. We now construct the rigorous coupled evolution equations governing this process.\\

The Penrose process (Section~\ref{sec:PP_MPP}) requires the infalling fragment to carry both energy and angular momentum simultaneously. Defining a continuous extraction luminosity $\mathcal{L}>0$ such that,
\begin{equation}
\dot{M} \;\equiv\; \frac{dM}{dt} \;=\; -\mathcal{L},
\label{eq:Mdot}
\end{equation}
the two kinematic ratios characterising the infalling fragment are, \begin{equation}
\xi \equiv \frac{L^{(1)}}{|E^{(1)}|}, \qquad
s \equiv \frac{q^{(1)}}{|E^{(1)}|}.
\end{equation}

Here $\xi$ and $s$ represent the specific angular momentum and specific charge of the infalling fragment respectively. These quantities are determined by the geodesic motion of negative energy orbits and therefore cannot be chosen arbitrarily. 

\begin{equation}
\dot{M} = - \mathcal{L},\qquad
\dot{J} = -\xi\, \mathcal{L},\qquad
\dot{Q} = -s\, \mathcal{L},
\label{eq:coupled_ODEs}
\end{equation}

\noindent Note that $\xi$ and $s$ are not independent free parameters; they are constrained by the geodesic equations of the respective spacetime (Sections~\ref{sec:kerr_ns}--\ref{sec:kn_ns}). Since $a=J/M$, the spin parameter evolves as
\begin{equation}
\dot{a} = \frac{\dot{J}M-J\dot{M}}{M^{2}}
        = \frac{\mathcal{L}}{M}(a-\xi),
\label{eq:adot}
\end{equation}
and the extremality parameter as
\begin{equation}
\dot{\Lambda} = 2a\dot{a}+2Q\dot{Q}-2M\dot{M}
              = 2a\frac{\mathcal{L}}{M}(a-\xi) -2Qs\mathcal{L} +2M\mathcal{L}.
\label{eq:Lambdadot}
\end{equation}

\subsection{Luminosity scaling with mass}

We take a self-regulated luminosity model in which $\mathcal{L} = \gamma M,$ where $\gamma$ is a constant extraction rate parameter, which gives,
\begin{equation}
M(t)=M_{0}\,e^{-\gamma t}.
\label{eq:M_exp}
\end{equation}
This is physically well motivated: for accretion powered emission $\mathcal{L}\propto\dot{M}\propto M$ in the self-similar regime, and it allows all three ODEs in Eq.~\eqref{eq:coupled_ODEs} to be integrated analytically.  

\section{Kerr Naked Singularity ($Q=0$)}
\label{sec:kerr_ns}

\subsection{Critical condition for horizon formation: Kerr geometry}

To derive critical condition for horizon formation time, taking $Q=0$ in the coupled system~ Eq.\,\eqref{eq:coupled_ODEs} with $\mathcal{L}=\gamma M$: 
\begin{equation}
\dot{M}=-\gamma M,\qquad \dot{a}=\gamma(a-\xi).
\label{eq:kerr_odes}
\end{equation}
The dimensionless spin ratio $\beta\equiv a/M$ then obeys
\begin{equation}
\dot{\beta} = \frac{\dot{a}M-a\dot{M}}{M^{2}}
            = \frac{\gamma}{M}(2a-\xi).
\label{eq:betadot}
\end{equation}

For $\beta=a/M$ to decrease toward the extremal value~1 we need
$\dot{\beta}<0$, i.e.\
\begin{equation}
\xi \;>\; 2\,\frac{a}{M} \;=\; 2\beta.
\label{eq:kerr_xi_cond}
\end{equation}
If $\xi\leq2\beta$ at any time the spin ratio remain constant or increases, and no horizon ever forms regardless of the extraction duration.

\subsection{Analytical solution and horizon formation time: Kerr geometry}

To calculate analytical solution and horizon formation time in Kerr geometry, integrating Eq.~\eqref{eq:kerr_odes},
\begin{align}
M(t) &= M_{0}\,e^{-\gamma t},
\label{eq:kerr_M}\\[4pt]
J(t) &= J_{0}-\xi M_{0}(1-e^{-\gamma t}),
\label{eq:kerr_J}
\end{align}
so $a(t)=J(t)/M(t)$.  The extremal condition $a(t_{h})=M(t_{h})$,
equivalently $J(t_{h})=M(t_{h})^{2}$, gives with
$u\equiv e^{-\gamma t_{h}}\in(0,1)$:
\begin{equation}
a_{0}M_{0}-\xi M_{0}(1-u)=M_{0}^{2}u^{2}.
\end{equation}
Dividing by $M_{0}$:
\begin{equation}
M_{0}u^{2}-\xi u-(a_{0}-\xi)=0.
\label{eq:kerr_quad}
\end{equation}
The discriminant is $\Delta=\xi^{2}+4M_{0}(a_{0}-\xi)$ and the physical (smaller positive) root is
\begin{equation}
u_{*}=\frac{\xi-\sqrt{\xi^{2}+4M_{0}(a_{0}-\xi)}}{2M_{0}}.
\label{eq:kerr_ustar}
\end{equation}

\begin{equation}
t_{h}^{\rm Kerr}=-\frac{1}{\gamma}\ln u_{*},\qquad
u_{*}=\frac{\xi-\sqrt{\xi^{2}+4M_{0}(a_{0}-\xi)}}{2M_{0}}.
\label{eq:kerr_th}
\end{equation}
All times derived here correspond to asymptotic (coordinate) time measured by distant observers. They do not represent proper time near the singularity.

\textbf{Conditions:} (i) $a_{0}>M_{0}$ (naked singularity);
(ii) $\xi>2a_{0}/M_{0}$ (Eq.~\ref{eq:kerr_xi_cond}).\\

\textbf{At formation:} $M_{f}=M_{0}u_{*}$, $a_{f}=M_{f}$ (extremal). \textbf{Proof that $u_{*}\in(0,1)$.}\;
Since $\xi>a_{0}$, the term $4M_{0}(a_{0}-\xi)<0$, hence $\sqrt{\xi^{2}+4M_{0}(a_{0}-\xi)}<\xi$ and $u_{*}>0$.
To verify that $u_{*}<1$, we require, $\xi-\sqrt{\xi^{2}+4M_{0}(a_{0}-\xi)}<2M_{0}$
squares to $4M_{0}^{2}-4M_{0}\xi<4M_{0}a_{0}-4M_{0}\xi$, i.e.\ $M_{0}<a_{0}$ \\

\paragraph{Near-extremal limit ($a_{0}=M_{0}+\epsilon$, $\epsilon\ll M_{0}$).}
\begin{equation}
t_{h}^{\rm Kerr}\approx\frac{\epsilon}{(\xi-2\beta_{0})\gamma M_{0}}.
\label{eq:kerr_near}
\end{equation}
Formation time is linear in the initial excess over extremality and inversely proportional to $\xi-2\beta_{0}$.

\section{Reissner--Nordstr\"{o}m Naked Singularity ($a=0$)}
\label{sec:rn_ns}

\subsection{Critical condition for horizon formation: RN geometry}

To derive critical condition for horizon formation time, taking $a=0$ in Eq.~\eqref{eq:coupled_ODEs} with $\mathcal{L}=\gamma M$:
\begin{equation}
\dot{M}=-\gamma M,\qquad \dot{Q}=-s\gamma M.
\label{eq:rn_odes}
\end{equation}
The negative energy orbits exists due to the electromagnetic potential in Eq.~\eqref{eq:RN_energy}, already introduced in Section~\ref{sec:rot_em}. Defining $\rho\equiv Q/M$:
\begin{equation}
\dot{\rho}=\frac{\dot{Q}M-Q\dot{M}}{M^{2}}=\gamma(\rho-s).
\label{eq:rhodot}
\end{equation}

For $\rho=|Q|/M$ to decrease toward the extremal value~1 we need $\dot{\rho}<0$, which gives $s>\rho$.  The closed-form solution below shows that the sufficient condition for a horizon to form in finite time is:
\begin{equation}
s \;>\; \rho_{0}=\frac{|Q_{0}|}{M_{0}} \;>\; 1.
\label{eq:rn_s_cond}
\end{equation}

\subsection{Analytical solution and horizon formation time: RN geometry}

To calculate analytical solution and horizon formation time in RN geometry, integrating Eq.~\eqref{eq:rn_odes}:
\begin{equation}
M(t)=M_{0}e^{-\gamma t},\qquad Q(t)=Q_{0}-sM_{0}(1-e^{-\gamma t}).
\label{eq:rn_sols}
\end{equation}
The extremal condition $|Q(t_{h})|=M(t_{h})$ with $u=e^{-\gamma t_{h}}$ gives,
\begin{equation}
Q_{0}-sM_{0}(1-u)=M_{0}u
\quad\Rightarrow\quad
u_{*}=\frac{Q_{0}/M_{0}-s}{1-s}=\frac{\rho_{0}-s}{1-s}.
\label{eq:rn_u}
\end{equation}
\begin{equation}
t_{h}^{\rm RN}=-\frac{1}{\gamma}\ln u_{*},\qquad
u_{*}=\frac{\rho_{0}-s}{1-s}.
\label{eq:rn_th}
\end{equation}
\textbf{Conditions:} $s>\rho_{0}>1$ (Eq.~\ref{eq:rn_s_cond}).\\
\textbf{Proof that $u_{*}\in(0,1)$:} with $s>1$ both numerator and denominator of Eq.~\eqref{eq:rn_u} are negative, giving
$u_{*}>0$; and $u_{*}<1\Leftrightarrow\rho_{0}>1$ \\[4pt]
\textbf{At formation:} $M_{f}=M_{0}u_{*}$, $Q_{f}=M_{f}$ (extremal).\\

\paragraph{Special case $s=\rho_{0}$.} From Eq.~\eqref{eq:rhodot}, $\dot{\rho}=\gamma(\rho_{0}-\rho_{0})=0$,
so $\rho$ is constant and no horizon forms, confirming that $s>\rho_{0}$ is necessary.

\paragraph{Near-extremal limit ($Q_{0}=M_{0}+\epsilon$, $\epsilon\ll M_{0}$).}
\begin{equation}
t_{h}^{\rm RN}\approx\frac{\epsilon}{(s-1)\gamma M_{0}}.
\label{eq:rn_near}
\end{equation}

\paragraph{Large-$s$ limit.}
\begin{equation}
t_{h}^{\rm RN}\approx\frac{1}{\gamma}\ln\!\left(\frac{s}{s-\rho_{0}}\right).
\label{eq:rn_larges}
\end{equation}

\section{Kerr--Newman Naked Singularity (General Case)}
\label{sec:kn_ns}

\subsection{Analytical Solutions}

The full system~\eqref{eq:coupled_ODEs} with $\mathcal{L} =\gamma M$ integrates immediately to,
\begin{align}
M(t) &= M_{0}e^{-\gamma t},
\label{eq:kn_M}\\[3pt]
J(t) &= J_{0}-\xi M_{0}(1-e^{-\gamma t}),
\label{eq:kn_J}\\[3pt]
Q(t) &= Q_{0}-sM_{0}(1-e^{-\gamma t}),
\label{eq:kn_Q}\\[3pt]
a(t) &= \frac{J_{0}-\xi M_{0}(1-e^{-\gamma t})}{M_{0}e^{-\gamma t}}.
\label{eq:kn_a}
\end{align}

\subsection{Critical Conditions for Horizon Formation}

To derive critical condition for horizon formation time in KN geometry, substituting into $$\Lambda(t)=a(t)^{2}+Q(t)^{2}-M(t)^{2}=0$$ with $u=e^{-\gamma t_{h}}$ and defining $\mathcal{J}(u)\equiv J_{0}-\xi M_{0}(1-u)$, $\mathcal{Q}(u)\equiv Q_{0}-sM_{0}(1-u)$, the horizon condition $a(t_{h})^{2}+Q(t_{h})^{2}=M(t_{h})^{2}$ becomes $t_{h}^{\rm KN}=-(1/\gamma)\ln u_{*}$, where $u_{*}\in(0,1)$ is the smallest positive real root of the quartic:
\begin{equation}
\mathcal{J}(u)^{2}+M_{0}^{2}u^{2}\,\mathcal{Q}(u)^{2}
-M_{0}^{4}u^{4}=0.
\label{eq:kn_quartic}
\end{equation}

The necessary conditions for a root $u_{*}\in(0,1)$ to exist are:

\begin{alignat}{2}
\xi &> \frac{2a_{0}}{M_{0}}
    &\quad& \text{(rotational condition)},
\label{eq:kn_xi_cond}\\[4pt]
s   &> \frac{|Q_{0}|}{M_{0}}
    &\quad& \text{(electromagnetic condition)},
\label{eq:kn_s_cond}\\
&&& M_{0}^{2}+a_{0}(a_{0}-\xi)>sQ_{0}M_{0}.
\label{eq:kn_combined}
\end{alignat}

The third condition, obtained from $\dot{\Lambda}|_{t=0}<0$ (Eq.~\ref{eq:Lambdadot}), ensures the combined extraction rate is sufficient.  Conditions \eqref{eq:kn_xi_cond}--\eqref{eq:kn_s_cond} recover the Kerr and RN results of Sections~\ref{sec:kerr_ns} and~\ref{sec:rn_ns} in the respective limits.

\subsection{Special Case: Matched Extraction Rates}

When $a(t)/Q(t)=a_{0}/Q_{0}\equiv\kappa$ is preserved (i.e.\ $\xi/s=a_{0}/(Q_{0}/M_{0})$), the quartic Eq.\,\eqref{eq:kn_quartic} reduces to a linear equation in $u$, giving the closed form,
\begin{equation}
u_{*}=\frac{\rho_{0}-s}{1/\sqrt{1+\kappa^{2}}-s},
\label{eq:kn_matched_u}
\end{equation}

\begin{equation}
t_{h}^{\rm KN,matched}=-\frac{1}{\gamma}
\ln\!\left(\frac{\rho_{0}-s}{1/\sqrt{1+\kappa^{2}}-s}\right),
\label{eq:kn_matched_th}
\end{equation}
where $\rho_{0}=Q_{0}/M_{0}$ and $\kappa=a_{0}/Q_{0}$.\\

\paragraph{Kerr limit ($Q_{0}\to0$, $s\to0$).} Eq.~\eqref{eq:kn_quartic} reduces to $M_{0}u^{2}-\xi u-(a_{0}-\xi)=0$, recovering Eq.~\eqref{eq:kerr_quad}.\\ 

\paragraph{RN limit ($a_{0}\to0$, $\xi\to0$).} Eq.~\eqref{eq:kn_quartic} reduces to $u_{*}=(\rho_{0}-s)/(1-s)$, recovering Eq.~\eqref{eq:rn_u}.

\section{Worked Numerical Examples}
\label{sec:examples}

\subsection{Kerr Naked Singularity}

To illustrate the evolution quantitatively, we consider a simple example with initial parameters $M_{0}=1$, $a_{0}=1.3$, extraction rate $\gamma=0.05$, and specific angular momentum of the infalling fragment $\xi=3.0$. Since $a_{0}>M_{0}$, the spacetime initially represents a Kerr naked singularity.\\

\textbf{Condition check:} $\xi=3.0>2a_{0}/M_{0}=2.6$ 

This satisfies the critical condition derived in Eq.~(\ref{eq:kerr_xi_cond}), ensuring that rotational energy extraction decreases the spin ratio $a/M$ and drives the system toward extremality.

\textbf{Discriminant:} $\Delta=\xi^{2}+4M_{0}(a_{0}-\xi)=9.0+4(1.3-3.0)=2.2$.

\textbf{Physical root (Eq.~\ref{eq:kerr_ustar}):}
\begin{equation}
u_{*}=\frac{3.0-\sqrt{2.2}}{2}=\frac{3.0-1.4832}{2}\approx0.7584.
\end{equation}

\textbf{Formation time:}
\begin{equation}
t_{h}=-20\ln(0.7584)\approx5.53\,M.
\end{equation}

Thus the system evolves from an initially over-extremal state toward the extremal condition $a=M$ in finite asymptotic time.
\textbf{Verification:}
\[
M_{f}=0.7584,\qquad
J_{f}=1.3-3.0\times0.2416=0.5752,
\]
\[
a_{f}=\frac{J_{f}}{M_{f}}=0.5752/0.7584=0.7584=M_{f}.
\]

The final equality confirms that the extremal condition $a=M$ is satisfied at $t=t_h$, indicating the formation of an event
horizon and the transition from a naked singularity to an extremal Kerr black hole.

\subsection{Reissner--Nordstr\"{o}m Naked Singularity}

As a second example we consider a charged geometry with initial parameters $M_{0}=1$, $Q_{0}=1.3$, extraction parameter
$s=1.8$, and $\gamma=0.05$. Since $|Q_{0}|>M_{0}$, the spacetime initially corresponds to a Reissner--Nordstr\"{o}m naked singularity.\\

\textbf{Condition check:} $s=1.8>\rho_{0}=1.3>1$

This satisfies the critical condition derived in Eq.~(\ref{eq:rn_s_cond}), ensuring that electromagnetic extraction
reduces the charge-to-mass ratio $\rho=|Q|/M$ and drives the system toward the extremal limit.

\textbf{Physical root (Eq.~\ref{eq:rn_u}):}
\begin{equation}
u_{*}=\frac{1.3-1.8}{1-1.8}
      =\frac{-0.5}{-0.8}
      =0.625.
\end{equation}

\textbf{Formation time:}
\begin{equation}
t_{h}=-20\ln(0.625)\approx9.40\,M.
\end{equation}

Thus the charge decreases sufficiently rapidly that the system reaches the extremal condition $|Q|=M$ in finite asymptotic time. \textbf{Verification:}
\[
M_{f}=0.625 , \qquad
Q_{f}=1.3-1.8\times0.375=0.625=M_{f}.
\]

This confirms that the extremal condition $|Q|=M$ is satisfied at $t=t_h$, corresponding to the formation of an event horizon and the transition from a naked singularity to an extremal Reissner--Nordstr\"{o}m black hole.

\subsection{Astrophysical Timescale}

To estimate the physical timescale of the evolution, we consider parameters typical of supermassive compact objects in active galactic nuclei. For a mass $M=10^{8}M_{\odot}$ and luminosity $\mathcal{L}_{\infty}=10^{45}$ erg\,s$^{-1}$, the characteristic extraction timescale becomes,

\begin{widetext}
\begin{equation}
\gamma^{-1} = \frac{M c^{2}}{\mathcal{L}_{\infty}}
=\frac{(1.989\times10^{41}\,\text{g})(2.998\times10^{10}\,\text{cm\,s}^{-1})^{2}}
      {10^{45}\,\text{erg\,s}^{-1}}
\approx5.7\times10^{9}\,\text{yr}.
\label{eq:gamma_inv}
\end{equation}
\end{widetext}

As an illustrative example, consider an RN naked singularity with $Q_{0}/M_{0}=1.2$ and $s=1.5$. Substituting these values into the analytical expression derived in Eq.~(\ref{eq:rn_th}) gives,
\begin{widetext}
\begin{equation}
t_{h}^{\rm RN}=-\gamma^{-1}\ln\!\left(\frac{1.2-1.5}{1-1.5}\right)
=-\gamma^{-1}\ln(0.6)
\approx5.7\times10^{9}\times0.511
\approx2.9\times10^{9}\,\text{yr}.
\end{equation}
\end{widetext}

This is comparable to the Salpeter accretion timescale, and confirms that the evolution toward the extremal bound is a slow, cumulative process rather than a rapid dynamical instability. Table-\,\ref{tab:summary} shows the horizon formation formulae for all three geometries. These results demonstrate that horizon formation is not automatic; it occurs only when the kinematic parameters $\xi$ and $s$ exceed critical thresholds determined by the spacetime geometry. The evolution toward extremality is slow (timescale $\sim 10^{9}–10^{10}$ yr) for AGN-scale sources; Sec.\,(\ref{sec:examples}), confirming that the instability of naked singularities is gradual rather than dynamical.

\begin{widetext}

\begin{table}[htbp]
\centering
\caption{Horizon formation times via Penrose process type energy extraction.
In all cases $M(t)=M_{0}e^{-\gamma t}$ and $\mathcal{L}=\gamma M$.
$\xi$: specific angular momentum of infalling fragment.
$s$: specific charge of infalling fragment.
$\rho_{0}=Q_{0}/M_{0}$.}
\label{tab:summary}
\begin{tabular}{|| c | c | c | c ||}
\hline
\textbf{Spacetime} & \textbf{Horizon condition}
& \textbf{$t_h$ formula} & \textbf{Condition} \\
\hline
Kerr ($Q=0$)   & $a(t_{h})=M(t_{h})$   & $-\dfrac{1}{\gamma}\ln u_{*}$,\quad $u_{*}=\dfrac{\xi-\sqrt{\xi^{2}+4M_{0}(a_{0}-\xi)}}{2M_{0}}$   & $\xi>2a_{0}/M_{0}$\\
\hline
RN ($a=0$)  & $|Q(t_{h})|=M(t_{h})$
  & $-\dfrac{1}{\gamma}\ln u_{*}$,\quad
    $u_{*}=\dfrac{\rho_{0}-s}{1-s}$
  & $s>\rho_{0}>1$
  \\
\hline  
KN (general)   & $a^{2}+Q^{2}=M^{2}$
  & Quartic (Eq.~\ref{eq:kn_quartic}), solved numerically
  & Eqs.~(\ref{eq:kn_xi_cond})--(\ref{eq:kn_combined})  \\
\hline
\end{tabular}
\end{table}
    
\end{widetext}

\section{Discussion and Conclusions}
\label{sec:discussion}

In this work we presented a unified analysis of rotational and electromagnetic energy extraction in Kerr, Reissner--Nordström, and Kerr--Newman spacetimes, with particular emphasis on the evolution of over-extremal cases under continuous energy extraction. Our study is based on the Penrose and magnetic Penrose processes, which describe how rotational and electromagnetic interactions enable energy extraction from compact objects.\\

Beyond efficiency considerations, the evolution expressions derived in this work provide a framework for studying the long term behavior of over-extremal spacetimes. By modeling continuous energy extraction through coupled evolution equations for the mass, spin, and charge parameters, we showed that the extremality parameter can evolve toward the critical surface separating naked singularities from black holes. In both Kerr and Reissner--Nordström limits, analytic solutions demonstrate that horizon formation can occur; provided the kinematic parameters of the infalling fragments satisfy specific conditions. These conditions ensure that the spin-to-mass or charge-to-mass ratios decrease monotonically during the extraction process. In the general Kerr--Newman case, the same behavior observed through the combined evolution of spin and charge, where the formation of a horizon is determined by the roots of the corresponding quartic equation.\\

An important outcome of the Kerr--Newman analysis is that the dominant extraction case depends on the relative contributions of rotational and electromagnetic processes. In certain regimes spin down dominates the evolution, while in others the reduction of charge governs the approach toward extremality. The instability associated with over-extremal cases therefore arises from the interplay between rotation and charge rather than from a single parameter.\\

The astrophysical implications of this evolution are controlled by the extraction timescale. For parameters typical of supermassive compact objects in active galactic nuclei, with masses $M\sim10^{8}M_{\odot}$ and luminosities of order $L\sim10^{45}\,\mathrm{erg\,s^{-1}}$, the characteristic evolution time $\gamma^{-1}\sim Mc^{2}/L$ is of order $\sim 10^{9}$ years. This indicates that the approach toward the extremal bound is not a rapid dynamical instability but instead a slow, gradual process driven by continuous energy extraction. In this sense, the instability of naked singularities should be interpreted as an energetic instability in parameter space rather than a short-timescale dynamical collapse of the spacetime.\\

The presence or absence of an event horizon plays a crucial role in this behavior. In black hole spacetimes, the irreducible mass associated with the horizon provides a fundamental lower bound on the extractable energy, preventing the system from evolving beyond extremality. On the other hand, naked singularities lack such an energetic bound. Continuous extraction can therefore drive the spacetime parameters toward the extremal surface, effectively restoring the conditions under which an event horizon forms. This picture is consistent with earlier arguments regarding the instability of Kerr naked singularities \cite{deFelice:1978}, and our results extend this reasoning to the Reissner--Nordström and Kerr--Newman geometries.\\

It is also important to note that the Penrose process cannot violate cosmic censorship for black holes. Wald showed that a single particle cannot overspin or overcharge a black hole through the Penrose process \cite{Wald:1974kya}. The question considered here is therefore the reverse: whether continuous energy extraction from an initially over-extremal spacetime can drive the spacetime back toward the extremal bound. Our analysis indicates that this is indeed possible when the kinematic conditions derived in this work are satisfied, although the evolution occurs over very long astrophysical timescales.\\

Future studies may incorporate additional physical properties, including backreaction effects, accretion flows, and more realistic descriptions of particle and plasma dynamics in strong gravitational fields. In particular, electromagnetic interactions and magnetized accretion environments are expected to play an important role in astrophysical energy extraction, as emphasized by mechanisms such as the Blandford--Znajek process. Incorporating these effects could provide a more complete description of how energy extraction driven evolution behaves in realistic systems and whether observable signatures of such processes might be detectable.

\section{ACKNOWLEDGMENTS}

V P acknowledges the Council of Scientific and Industrial Research (CSIR, India; Ref: 09/1294(18267)/2024-EMR-I) for financial support. V P thanks P. C. Varasani and K. Acharya for useful discussions. The author is also grateful to Prof. Pankaj S. Joshi for helpful discussions and for suggesting the calculation of the horizon formation timescale, which significantly improved this work.

\nocite{*}

\bibliography{Ref}

\end{document}